\documentclass[useAMS,usenatbib]{mn2e}

\usepackage{graphicx}
\usepackage{scalefnt}
\usepackage{amssymb}
\usepackage{upgreek}

\def\etal{et al.~}  
\def\ie{{i.e.~}}
\def\galform{\texttt{GALFORM}}
\def\grasil{\texttt{GRASIL}}

\newcommand{\ergsec}	{\ifmmode \mathrm{\,erg~s}^{-1} \else \,erg~s$^{-1}$\fi}
\newcommand{\yr}	    {\ifmmode \mathrm{yr} \else yr\fi}
\newcommand{\mpc}	{\ifmmode \,\mathrm{Mpc}^{-3} \else \,Mpc$^{-3}$\fi}
\newcommand{\Msun}	{\ifmmode \,\mathrm M_{\odot} \else $\,\mathrm M_{\odot}$\fi}
\newcommand{\Mbh}	{\ifmmode M_{\mathrm{BH}} \else $M_{\mathrm{BH}}$\fi}
\newcommand{\Mseed}	{\ifmmode M_{\mathrm{BH,seed}} \else $M_{\mathrm{BH,seed}}$\fi}
\newcommand{\Mbulge}	{\ifmmode M_{\mathrm{Bulge}} \else $M_{\mathrm{Bulge}}$\fi}
\newcommand{\Mhalo}	{\ifmmode M_{\mathrm{Halo}} \else $M_{\mathrm{Halo}}$\fi}
\newcommand{\Mhaloeff}	{\ifmmode M_{\mathrm{Halo,eff}} \else $M_{\mathrm{Halo,eff}}$\fi}
\newcommand{\Medd}	{\ifmmode \dot{M}_{\mathrm{Edd}} \else
v  $\dot{M}_{\mathrm{Edd}}$\fi}
\newcommand{\Mstar}	{\ifmmode \dot{M}_{\mathrm{*}} \else $\dot{M}_{\mathrm{*}}$\fi}
\newcommand{\Lbol}	{\ifmmode L_{\mathrm{bol}} \else
  $L_{\mathrm{bol}}$\fi}
\newcommand{\Lx}	    {\ifmmode L_{\mathrm{X}} \else $L_{\mathrm{X}}$\fi}
\newcommand{\vlv}	{\ifmmode \nu L_{\mathrm{\nu}} \else $\nu L_{\mathrm{\nu}}$\fi}
\newcommand{\Lagn}	{\ifmmode L_{\mathrm{AGN}} \else $L_{\mathrm{AGN}}$\fi}
\newcommand{\Ledd}	{\ifmmode L_{\mathrm{Edd}} \else $L_{\mathrm{Edd}}$\fi}
\newcommand{\Lcool}	{\ifmmode L_{\mathrm{cool}} \else $L_{\mathrm{cool}}$\fi}
\newcommand{\Lsun}	{\ifmmode L_{\odot} \else $L_{\odot}$\fi}
\newcommand{\Lsx}	{\ifmmode L_{\mathrm{SX}} \else $L_{\mathrm{SX}}$\fi}
\newcommand{\Mbj}	{\ifmmode M_{b_{\rm J}} \else $M_{b_{\rm J}}$\fi}
\newcommand{\bj}		{\ifmmode b_{\rm J} \else $b_{\rm J}$\fi}
\newcommand{\Lhx}	{\ifmmode L_{\mathrm{HX}} \else $L_{\mathrm{HX}}$\fi}
\newcommand{\ledd}	{\ifmmode \lambda_{\mathrm{Edd}} \else $\lambda_{\mathrm{Edd}}~$\fi}
\newcommand{\lgledd}	{\ifmmode \lambda_{\mathrm{Edd}} \else $\lambda_{\mathrm{Edd}}$\fi}
\newcommand{\rhobh}	{\ifmmode \rho_{\mathrm{BH}} \else $\rho_{\mathrm{BH}}$\fi}
\newcommand{\mdot}	{\ifmmode \dot{m} \else $\dot{m}$\fi}
\newcommand{\fhalo}	{\ifmmode f_{\mathrm{Halo}}^{\mathrm{act}} \else $f_{\mathrm{Halo}}^{\mathrm{act}}$\fi}
\newcommand{\fvis}	{\ifmmode f_{\mathrm{vis}} \else $f_{\mathrm{vis}}$\fi}
\newcommand{\fobsc}	{\ifmmode f_{\mathrm{obsc}} \else $f_{\mathrm{obsc}}$\fi}
\newcommand{\fq}		{\ifmmode f_{\mathrm{q}} \else $f_{\mathrm{q}}$\fi}
\newcommand{\fbh}	{\ifmmode F_{\mathrm{BH}} \else $F_{\mathrm{BH}}$\fi}
\newcommand{\Lir}	{\ifmmode L_{\mathrm{IR}} \else $L_{\mathrm{IR}}$\fi}
\newcommand{\Lsixty}	{\ifmmode \nu L_{60\mathrm{\mu m}} \else $\nu L_{60\mathrm{\mu m}}$\fi}

\title[SFR- AGN luminosity correlation]{The Star Formation and AGN luminosity
  relation: Predictions from a semi-analytical model}
\author[T. A. Gutcke, N. Fanidakis, A.V. Macci\`o, \& C. Lacey]{Thales A. Gutcke$^1$\thanks{thales@mpia.de}, Nikos Fanidakis$^1$, Andrea V. Macci\`o$^1$, \& Cedric Lacey$^2$\\
1 Max-Planck-Institut f\"ur Astronomie, K\"onigstuhl 17, 69117 Heidelberg, Germany\\
2 Institute for Computational Cosmology, Department of Physics, University of Durham, South Road, Durham, DH1 3LE, UK\\}
\begin{document}

\pagerange{\pageref{firstpage}--\pageref{lastpage}} \pubyear{---}

\maketitle

\label{firstpage}

\begin{abstract}

In a universe where AGN feedback regulates star formation in massive galaxies, a strong correlation between these two quantities is expected. If the gas causing star formation is also responsible for feeding the central black hole, then a positive correlation is expected. If powerful AGNs are responsible for the star formation quenching, then a negative correlation is expected.
Observations so far have mainly found a mild correlation or no correlation at all (\ie a flat
relation between star formation rate (SFR) and AGN luminosity), raising questions about the whole paradigm of ``AGN feedback''.
In this paper, we report the predictions of the $\galform$ semi-analytical model, which has a very strong coupling between AGN activity and quenching of star formation. The predicted SFR-AGN luminosity correlation appears negative in the low AGN luminosity regime, where AGN feedback acts, but becomes strongly positive in the regime of the brightest AGN. 
Our predictions reproduce reasonably well recent observations by \citeauthor{Rosario2012}, yet there is some discrepancy in the normalisation of the correlation at low luminosities and high redshifts. Though this regime could be strongly influenced by observational biases, we argue that the disagreement could be ascribed to the fact that $\galform$ neglects AGN variability effects.
Interestingly, the galaxies that dominate the regime where the observations imply a weak correlation are massive early-type galaxies that are subject to AGN feedback. Nevertheless, these galaxies retain high enough molecular hydrogen contents to maintain relatively high star formation rates and strong infrared emission.

\end{abstract}


\section{Introduction}

A clear understanding of how galaxies transition from star-forming disks to passive
spheroidals is one of the open problems in the current paradigm of galaxy formation.
One of the most widely  accepted theories requires a large energy injection
into the cores of massive galaxies; such energy will then
heat up the cold gas in the center and prevent the accretion of new
gas onto the galaxy. The most likely source for such energy are active
galactic nuclei (AGN), powered by matter accretion onto super massive
black holes (SMBHs) at the centers of galaxies \citep{Silk1998}.

{In the last decade both semi-analytical models and numerical simulations have tried to incorporate such an effect into the modelling of galaxy formation. Semi-analytical models have showed that if the energy from AGN is coupled to the hot and cold gas, it is indeed possible to halt star formation and create passive galaxies \citep{Kauffmann2000,Cattaneo2001, Granato2004, Bower2006,Croton2006, Somerville2008,Fanidakis2011, Guo2011}.
Numerical simulations have also helped to shed light onto the interplay between a galaxy and its SMBH (e.g.  \citealt{SpDMH2005,DiMatteo2005,Robertson2006,Sijacki2007,Hopkins2007,DiMatteo2008,Okamoto2008,Booth2009}. More recently, large fully cosmological simulations including the effect of AGN feedback have been quite successful in reproducing properties of observed galaxies, including the luminosity function and star formation rate (\citealt{Vogelsberger2014} and \citealt{Schaye2015}). There is general agreement that AGN feedback is a needed ingredient in galaxy formation and evolution.}


These studies imply a strong correlation between the evolution
of a galaxy and its central black hole. A hint to such a correlation
might reside in the so called M-sigma relation, namely the observed correlation between
the velocity dispersion of the bulge of the galaxy and the mass of the supermassive
black hole hosted by it (\citealt{Ferrarese2000};
\citealt{Gebhardt2000}, but see \citealt{Jahnke2011} for a different explanation).

Another possible piece of evidence of a causal link between the rising of AGN
activity and the quenching of a galaxy can be obtained by
studying the observed correlation between 
star formation rate (SFR) and AGN luminosity (\citealt{Shao2010},
\citealt{Lutz2010} and
\citealt{Harrison2012}).

Recently,  \citealt{Rosario2012} used  deep
infrared  and  X-ray  observations   in  the  COSMOS,
GOODS-NORTH  and  GOODS-SOUTH  fields to estimate  the  SFR  and  AGN
luminosity, respectively. They found a weak correlation
between SFR and AGN activity at all redshifts, suggesting little
connection between SF and BH growth in these systems, especially at
high redshift and low luminosity. \citealt{Mullaney2012a}  and \citealt{Stanley2015} find
a similar absence of correlation {(but see \citealt{Mullaney2012b} and \citealt{Rodighiero2015} for a possible correlation of SMBH growth rate and SFR). }

This quite unexpected lack of correlation between SFR and AGN luminosity has
been used to put constraints on the triggering of AGN activity
and black hole growth (\citealt{Neistein2014}), possibly suggesting
strong variability in AGN activity on time scales shorter than those
typical of star formation, which are of the order 100 Myrs
(\citealt{Hickox2014}).

In this paper, we take a deeper look into the expected correlation
between SFR and AGN activity. We present predictions from the $\galform$
semi-analytical  model  of  galaxy formation  (\citealt{Cole2000})  in
which AGN feedback is the key ingredient to explain the fading of star
formation in massive galaxies. $\galform$  has  been successful  in
reproducing a number of fundamental relations of galaxy evolution and
structure:
the luminosity and stellar mass functions of galaxies, the AGN
diversity and evolution, the evolution of Lyman-$\alpha$ emitters (LAEs) and Lyman-break galaxies (LBGs), the number
counts of sub-millimeter galaxies (SMGs), as well as the HI
and HII mass functions (\citealt{Baugh2005},\citealt{Bower2006}, \citealt{Fanidakis2011}, \citealt{Lacey2011} and \citealt{Lagos2012}).

Using the version of $\galform$ presented in  \citealt{Fanidakis2012} and
assuming full co-evolution
between the galaxy and its central  BH, we study the predicted
correlation  between  SF   and  AGN  luminosity  and   compare  it  to
observations and previous models of this relation. We stress that this correlation is not used as a constraint on the model, but presents a pure prediction of a model constrained by other observations.

The paper  is organised  as follows.  In Section  1, we  introduce the
$\galform$ semi-analytical model and describe the most important model
ingredients that are relevant for this study. In Section 2, we present
the predictions of  the model for the FIR luminosity  as a function of
AGN activity and compare to the  observations. In Section 3 we look at
the molecular gas  content of our galaxies to  understand the material
that is causing star formation. Finally, in Section 4, we discuss our
findings and present our conclusions.

\section{The model}
\label{sec:model}

Semi-analytic modelling in galaxy formation is used to simulate large numbers of galaxies in a computationally efficient manner, thus enabling statistical insight into the predictions of different galaxy formation models. It uses our understanding of cosmological structure formation in N-body simulations and adds a set of coupled differential equations to describe the physical processes of galaxy formation and evolution, making possible the analysis of a wide variety of galaxy and BH properties. $\galform$, one of the most extensively applied semi-analytical models, has been improved upon over time (\citealt{Cole2000}; \citealt{Baugh2005}; \citealt{Bower2006}; \citealt{Fanidakis2011}; \citealt{Lagos2011}) to include processes and adjustments that describe more accurately the formation and evolution of galaxies. Among the most important physical processes that are modelled in $\galform$ are: i) the formation and evolution of DM haloes in the $\Lambda$CDM cosmology, ii) gas cooling and disk formation in DM haloes, iii) star formation, supernovae feedback and chemical enrichment, iv) BH growth and AGN feedback, v) and the formation of bulges through galactic disk instabilities and galaxy mergers. 

The model is successful in reproducing many observations, including the luminosity and stellar mass functions of galaxies, the number counts of submillimeter galaxies, the evolution of Ly$\alpha$ and Ly-break galaxies, H and H$_{2}$ mass functions and the AGN diversity and evolution (\citealt{Baugh2005}; \citealt{Bower2006}; \citealt{Orsi2008}; \citealt{Kim2011}; \citealt{Lacey2011}; \citealt{Lagos2011}; \citealt{Lagos2012}; \citealt{Lagos2014a}; \citealt{Lagos2014b}; \citealt{Fanidakis2011}; \citealt{Fanidakis2012}; \citealt{Fanidakis2013a}; \citealt{Fanidakis2013b}; \citealt{Gonzalez-Perez2013}; \citealt{Gonzalez-Perez2014}). The model provides a very good fit to the HERSCHEL far-IR galaxy luminosity functions (LFs) and number counts (Lacey et al. in prep.; see also \citealt{Lacey2010}) and X-ray LFs of AGN in the $z=0-3$ universe (\citealt{Fanidakis2012}; Fanidakis \etal in prep.) and therefore it is an ideal tool for studying the correlation between SFR and AGN activity. 

For the purposes of this analysis, we use a version of the \citet{Fanidakis2012} model, updated to the cosmological parameters estimated by the $7$-year data release of WMAP (\citealt{Komatsu2011} and the Lacey et al. in prep. \galform~model). The fundamental predictions of the model for the AGN properties are described in \citet[see also Fanidakis \etal in prep.]{Fanidakis2013a, Fanidakis2013b}. The predictions presented in this analysis are calculated using merger trees extracted from the DM only N-body simulation Millennium WMAP-7 (Lacey et al. in prep.). The Millennium WMAP-7 simulation has almost the same mass resolution ($8.61\times10^8\Msun$), particle number ($10^{7}$) and box size ($500\,h^{-1}$\,Mpc) as the Millennium simulation \citep{Springel2005} and differs only in the background cosmology (which is in agreement with WMAP7 results).  We now describe in the rest of this section the main model processes and ingredients that are essential for understanding the $\galform$ predictions for the SFR and AGN activity.

\subsection{The two regimes of black holes growth}
\label{sec:twomodes}
$\galform$ assumes two regimes of black hole (BH) growth in galaxies. The first one is the ``starburst" regime, where a galaxy experiences a starburst and AGN activity with high star formation and BH accretion rates. The high efficiency both in SF and BH growth of this regime is due to efficient gas cooling in DM haloes with masses lower than $5\times10^{12}\Msun$. The onset of the starburst mode is triggered by either a merger or by a disk instability. In this regime, a fraction of the cold gas that turns into stars is accreted directly onto the BH. When only this regime is active, we expect a positive correlation between star formation rate (SFR) and active galactic nucleus (AGN) luminosity. 

In the ``hot-halo'' regime, the growth of BHs is tightly linked to the AGN feedback mechanism and the suppression of gas cooling in haloes typically more massive than $5\times10^{12}\Msun$. The gas feeding the BH during this mode is assumed to originate directly from the hot halo around the galaxy. The resulting accretion power is then coupled via a jet to the thermodynamical properties of the gas in the host halo and suppresses cooling, if the available heating power exerts the cooling luminosity of the gas, $L_{\rm cool}$. Galaxies that are subject to AGN feedback still exhibit some SF, due to remaining cold gas and new gas brought in by mergers, but it is expected to be lower than in actively SF galaxies. In this regime, we expect a negative correlation between SFR and (AGN) luminosity.

In addition to the starburst and hot-halo modes of BH growth, BH-BH mergers during galaxy encounters also contribute to the growth of BHs. However, this growth mode only redistributes the BH mass and does not add new baryons to the BHs.

\subsection{Bolometric accretion luminosity}

The gas accreted during the starburst mode is converted into an accretion rate, $\dot{M}$, by assuming that the accretion duration is proportional to the dynamical timescale of the host spheroid,
{\begin{equation}
\dot{M}=\frac{M_{\rm acc}}{f_{\rm q}t_{bulge}}.
\end{equation}
Here $M_{\rm acc}$ is fixed for every galaxy to $0.5$ percent of the mass that turns into stars during a starburst, $t_{\rm dyn}$ is the dynamical timescale of the host spheroid and $f_{\rm q}$ is a proportionality factor set to $10$ in \citet{Fanidakis2012}.}
In contrast, in the hot-halo mode, the accretion rate onto the BH is calculated directly from the cooling properties of the host DM halo, \ie
\begin{equation}
\dot{M}=\frac{L_{\rm cool}}{\epsilon_{\rm kin} c^2},
\end{equation}
where $\epsilon_{\rm kin}$ is the average kinetic efficiency of the jet during the AGN feedback. {$L_{\rm cool}$ is the quasi-hydrostatic cooling luminosity of the halo. This is chosen assuming that the flow will balance heating and cooling in the hot halo mode (i.e. if the Eddington ratio is sufficiently large.)} This accretion mode is associated with early type galaxies in massive haloes, with relatively low SFRs. Since the accretion process is responsible for shutting down SF in the host galaxy, a negative correlation between AGN luminosity and SFR is expected in this mode. 

The bolometric luminosity of the accretion flow, \Lbol, is calculated by assuming the Shakura-Sunyaev thin disk model \citep{Shakura1973}:
\begin{equation}
L_{\mathrm{bol,TD}} = \epsilon\dot{M}c^2.
\end{equation}
We assume that this solution is valid for accretion rates higher than $1$ percent of the Eddington accretion rate, \ie $\dot{m}=\dot{M}/\dot{M}_{\rm Edd}\geqslant 0.01$, where $\dot{M}_{\rm Edd}$ is defined as $\Ledd/\epsilon c^2$. For lower accretion rates, the advection-dominated accretion flow (ADAF) thick-disk solution is used (\citealt{Narayan1994}; \citealt{Mahadevan1997}),
\begin{eqnarray}
L_{\mathrm{bol,ADAF}}=0.44\left(\frac{\dot{m}}{0.01}\right)\epsilon\dot{M}c^2.
\label{disk_bol_adaf}
\end{eqnarray}
When the accretion becomes substantially super-Eddington ($\Lbol\geqslant\eta\Ledd$), the bolometric luminosity is limited to \citep{Shakura1973}:
\begin{equation}
\Lbol(\geqslant\eta\Ledd)=\eta[1+\ln(\dot{m}/\eta)]L_{\mathrm{Edd}},
\end{equation}
where $\eta$ is an ad hoc parameter, which we choose equal to $4$, to allow a better modelling of the bright end of the AGN luminosity function \citep[see][]{Fanidakis2012}. However, the accretion rate, \mdot, is not restricted if the flow becomes super-Eddington. 

\subsection{Star formation rate}\label{sfr_laws}

\citet{Lagos2011} recently revised the original formulation of the SF law
in $\galform$ \citep{Cole2000}, in favour of the observationally
motivated (and constrained) Blitz \& Rosolowsky (\citealt{Blitz2006}) empirical law. The Blitz \& Rosolowsky law is motivated by UV, FIR and millimetre observations of SF and molecular gas in spiral galaxies and assumes that the surface density of star formation rate (SFR) is proportional to the surface density of molecular hydrogen in the disk. 
\begin{equation}
  \label{eq:BR06-1}
  \Sigma_{\rm SFR} = \nu_{\rm SF}\Sigma_{\rm mol}.
\end{equation}
The proportionality factor, $\nu_{\rm SF}$, is given as an inverse time scale
and its value is closely constrained by observations. At every timestep, $\galform$ calculates the molecular-to-atomic hydrogen ratio in the disk, $R_{\rm mol}$,
which is expressed as a power law of the internal hydrostatic pressure of the
disk \citep{Elmegreen1993}:
\begin{equation}
R_{\rm mol} = {\rm \frac{\Sigma(H_{2})}{\Sigma(H)}}=\left(\frac{P_{\rm ext}}{P_{\rm 0}}\right)^{\alpha}.
\end{equation}
Eq. \ref{eq:BR06-1} is then re-written in terms of the cold gas surface density, $\Sigma_{\rm gas}$, as:
\begin{equation}
  \label{eq:BR06-2}
  \Sigma_{\rm SFR} = \nu_{\rm SF}f_{\rm mol}\Sigma_{\rm gas},
\end{equation}
where $f_{\rm mol} = R_{\rm mol}/(1+R_{\rm mol})$. Integration of Eqn.~\ref{eq:BR06-2} over the assumed exponential surface density profile of the gas gives the SFR of the galaxy.

When the galaxy experiences a burst of SF, triggered either by a
galaxy-galaxy merger or a disk instability, the model assumes that the entire
cold gas reservoir of the galaxy (atomic and molecular) is converted into
stars. The SF timescale in starbursts is finite and proportional to the
dynamical timescale of the host bulge. At a given time, we estimate the total
SFR in the galaxy as the sum of the quiescent SFR and burst SFR. 

\subsection{Stellar and dust emission}
\label{sec:seds}

$\galform$ includes a self-consistent model for calculating the stellar emission in every galaxy and its absorption and re-emission in the mid and far-IR ($8-1000\micron$) and sub-mm wavelengths by dust. The model is similar in outline to the spectro-photometric model $\grasil$ \citep{Silva1998}, with some simplifications that speed-up significantly the calculation. The stellar emission in the model is calculated based on the stellar population synthesis model of \citet{Bressan1998} and assuming that the stars have an axisymmetric distribution in the disk and bulge. Dust in the galaxies is assumed to be a two-phase medium, \ie diffuse low-density dust in the ISM and dense dust clouds enshrouding star forming regions. The amount of dust in each galaxy is determined by the total mass and metallicity of cold gas. The attenuation of stellar light by the dust is computed by interpolating the tabulated radiative transfer models of \citealt{Ferrara1999}. The FIR emission from the dust is calculated assuming that the cloud and diffuse dust components re-emit the absorbed starlight (\ie the difference between the stellar and dust attenuated SED, integrated over wavelength) as a modified blackbody. This constrains the dust temperature (constant within a galaxy and different for each component), which is then used to calculate the SED of each dust component. The total dust SED is then found by summing over the SEDs of the two components. This technique works well for wavelengths greater than $\sim 60\micron$.

The composite SED of stellar and dust emission of galaxies in $\galform$ provides a very good fit to galaxy number counts and luminosity function in the FIR (see Lacey et al. in prep.). Here we are interested in comparing the predictions of the model directly to far-IR observations of AGN with the Herschel/PACS instrument, at a mean rest-frame wavelength of $60\rm\mu m$. For model galaxies predicted by $\galform$, the monochromatic luminosity $\Lsixty = \vlv(60\mu\rm m)$ is typically $\sim0.65\Lir$, where $\Lir$ is the total FIR luminosity, defined as the integrated luminosity between 8 and 1000~$\micron$. This wavelength range includes the entire emission from dust, while excluding the intrinsic stellar emission. In the rest of the analysis, the $\galform$ predictions will be presented at $60\rm\mu m$, unless the comparison with observational data requires a different band or the total dust emission. We note that the galaxy SED predicted by $\galform$ does not include any AGN emission. Therefore, the dust emission predicted by the model is purely reprocessed stellar emission.

\begin{figure}
  \centering
 \includegraphics[width=0.47\textwidth]{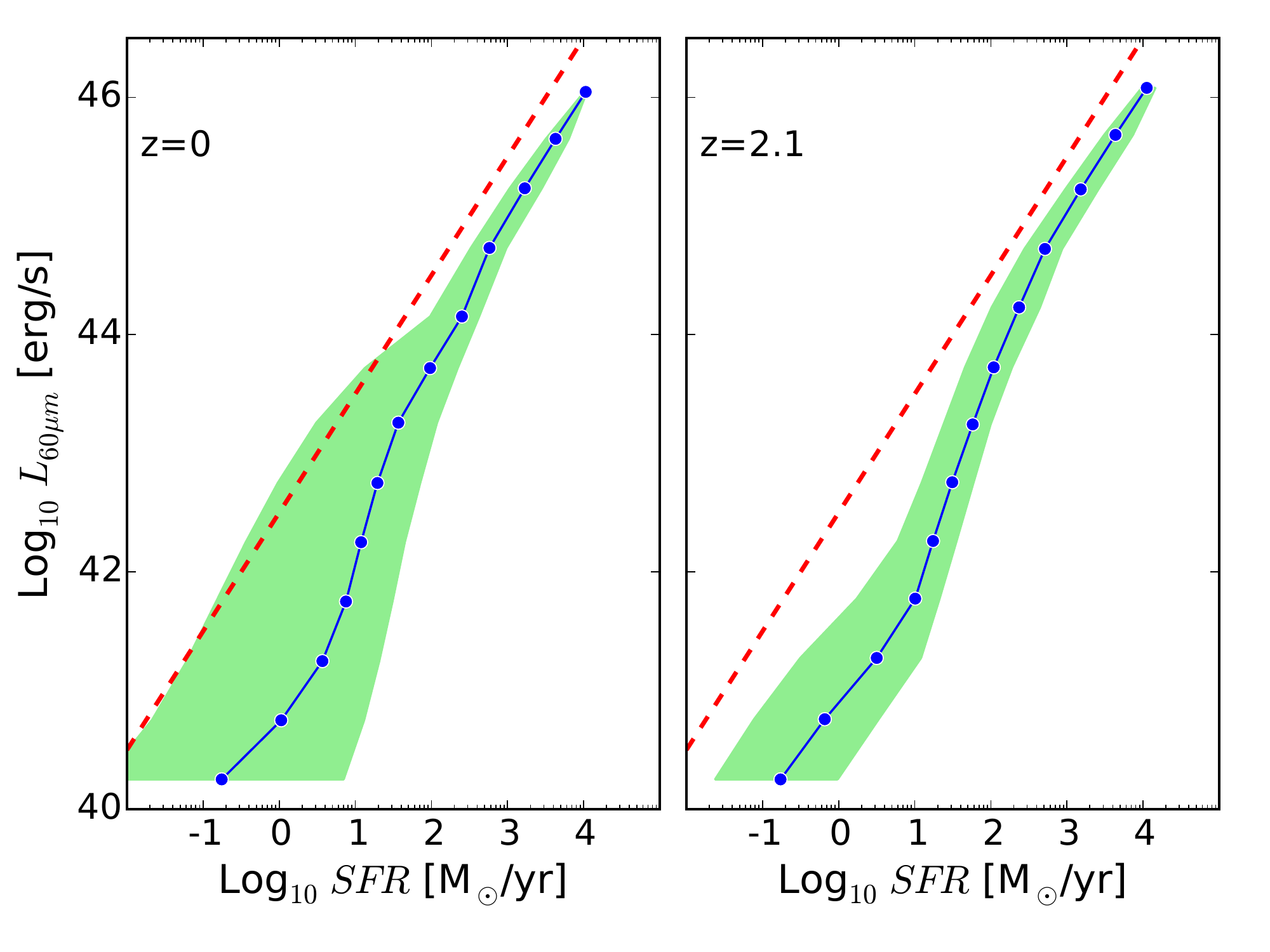}
  \caption{Predicted FIR emission at $60\;\micron$ as a function of
    star formation rate (SFR) at $z=0$ and $z=2.1$. The solid lines
    show median values, and the shaded areas the 10th and 90th
    percentiles. The red dashed line shows a line with a slope of unity and an arbitrary normalization.}
  \label{fig:sfr_Lir}
\end{figure}

Finally, we note that the SFR is usually traced by emission in the FIR as the SFR is not a direct observable. As a first test, we check how good of a proxy this emission is for the actual SFR in our model. Fig. \ref{fig:sfr_Lir} shows the median IR luminosity at $60\micron$ (blue line) and its 10th and 90th percentile (green shaded areas) as a function of SFR at $z=0$ and $2.1$. For high redshifts, the correlation is tighter, but even at $z=0$ it is still present over several orders of magnitude. 
The positive dependance of $\Lir$ on SFR is a consequence of the higher UV luminosity caused by the young stars. This in turn increases the dust heating and IR emission. At high SFRs, the $60 \mu m$ emission seems to be directly proportional to the SFR. At intermediate and low SFRs (${\rm SFR}\lesssim 10^1-10^2\; M_{\odot} yr^{-1}$) the correlation deviates mildly from a slope of unity mainly due to a somewhat lower dust extinction or lower dust temperature. Overall, we can conclude that the FIR emission strongly couples to the SFR, thus, IR luminosity can be used as a proxy for SFR. We will henceforth use the FIR emission at $\Lsixty$ as a proxy of star formation (SF).

\section{The FIR$-$AGN luminosity correlation}
\label{sec:results}

In this study we are interested in comparing the predicted and observed FIR$-$AGN luminosity correlations. Our final aim is to check what correlation of observable quantities is predicted by a model where AGN feedback is the driving mechanism for quenching star formation in high mass galaxies. 

\subsection{Model Predictions}

\begin{figure}
  \centering
  \includegraphics[width=0.47\textwidth]{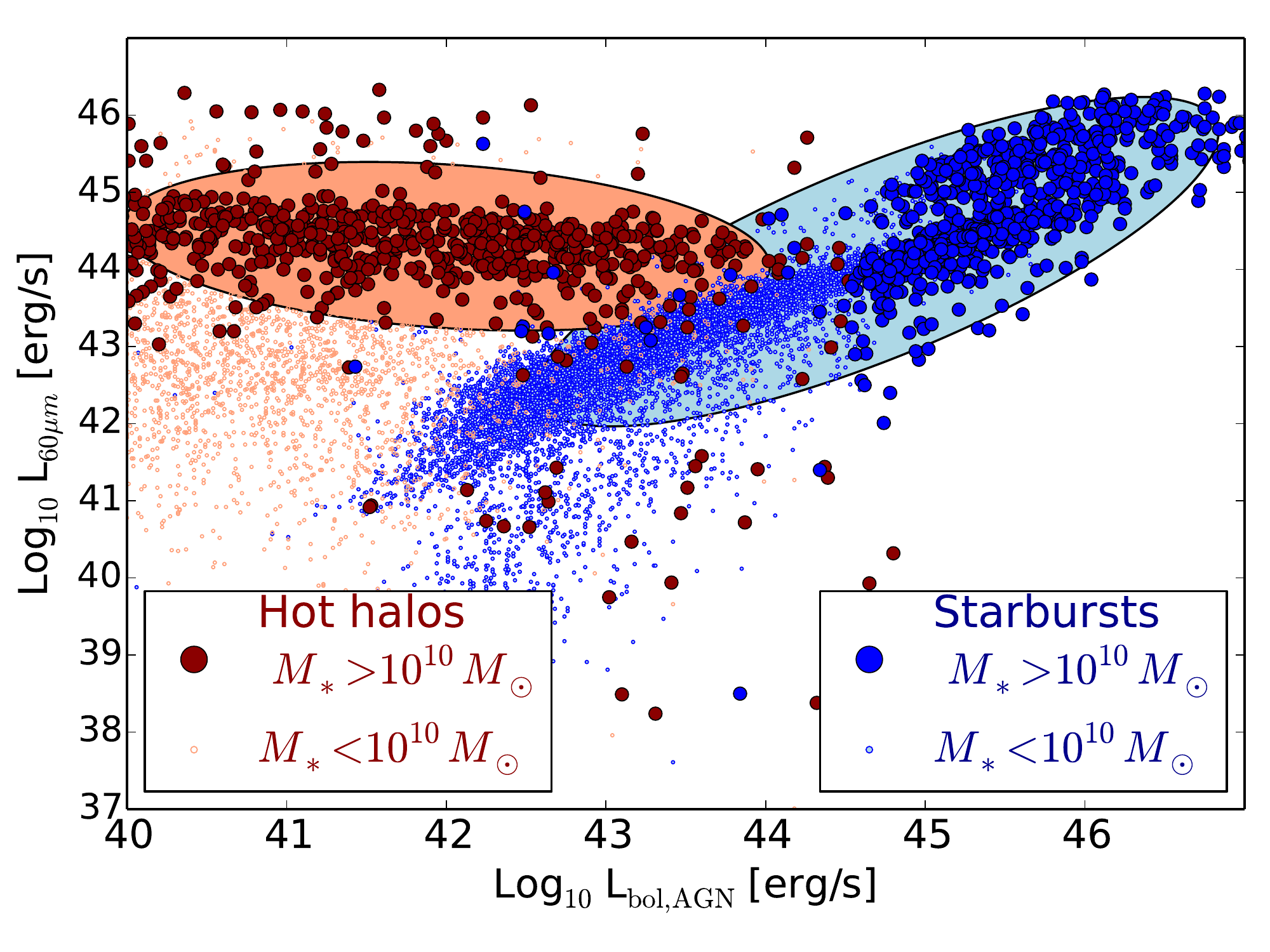}
  \caption{Predicted FIR emission at $60\;\micron$, $\Lsixty$, as a function of bolometric AGN luminosity, $\Lbol$, for $4,000$ randomly selected galaxies at $z=2.1$ (both centrals and satellites). Galaxies in which the central BH experiences starburst-mode activity are shown with blue, while those with BHs growing during the hot-halo mode with red. The point size indicates stellar masses below (small) and above (big) $10^{10}\Msun$.}
  \label{fig:scatter_plot}
\end{figure}

We show our first predictions for the FIR$-$AGN luminosity correlation in Fig.~\ref{fig:scatter_plot}, where we plot the distribution of galaxies on the $\Lbol-\Lsixty$ plane. These galaxies represent a small subset ($4,000$) of the total sample (usually of the order of $10^6$ galaxies) and are randomly selected from the model output at $z=2.1$. We split our sample into galaxies undergoing active starburst AGN activity (cold-gas accretion, blue points) and galaxies whose BHs are in the hot halo mode (hot-gas accretion, red points). We also divide the sample by stellar mass, with galaxies of $M_{*}>10^{10} M_{\odot}$ marked by a larger symbol, to aid the reader in finding the trends in galaxies that are more likely to be observed. Interestingly, low luminosity objects accreting in the hot-halo mode show a negative trend of $\Lsixty$ with AGN luminosity. In contrast, high luminosity AGN, powered through the starburst mode, show a clear positive trend. The two different trends are a manifestation of the fact that each of the two accretion modes is linked to a different regime of SF efficiency.

The different trends are due to the two different modes in which a BH accretes gas. Large AGN luminosities are due to accretion of cold gas and mostly happen in haloes with masses below $\sim5\times 10^{12} \Msun$, as shown in \citet{Fanidakis2013b}. In this case, the triggering mechanism (galaxy mergers or disk instabilities) of cold-gas flows that feed the central BH is also responsible for a burst of SF in the host galaxy, hence the positive correlation. Low luminosity AGN mainly live in high mass haloes ($\gtrsim 10^{13} \Msun$), where the gas fuelling the BH is accreted directly from the hot halo around the host galaxy (see Section \ref{sec:model}). Accretion activity in this mode is tightly linked to the AGN feedback mechanism and, thus, is responsible for the suppression of gas cooling and SF. The quenching nature of this mode gives rise to a negative correlation. In summary, the intrinsic $\Lsixty$ vs $\Lagn$ correlation predicted by $\galform$ is entirely shaped by the physics of each of the accretion modes that are responsible for growing the central BH. 

\subsection{Comparison with Data}

\begin{figure}
  \centering
 \includegraphics[width=0.47\textwidth]{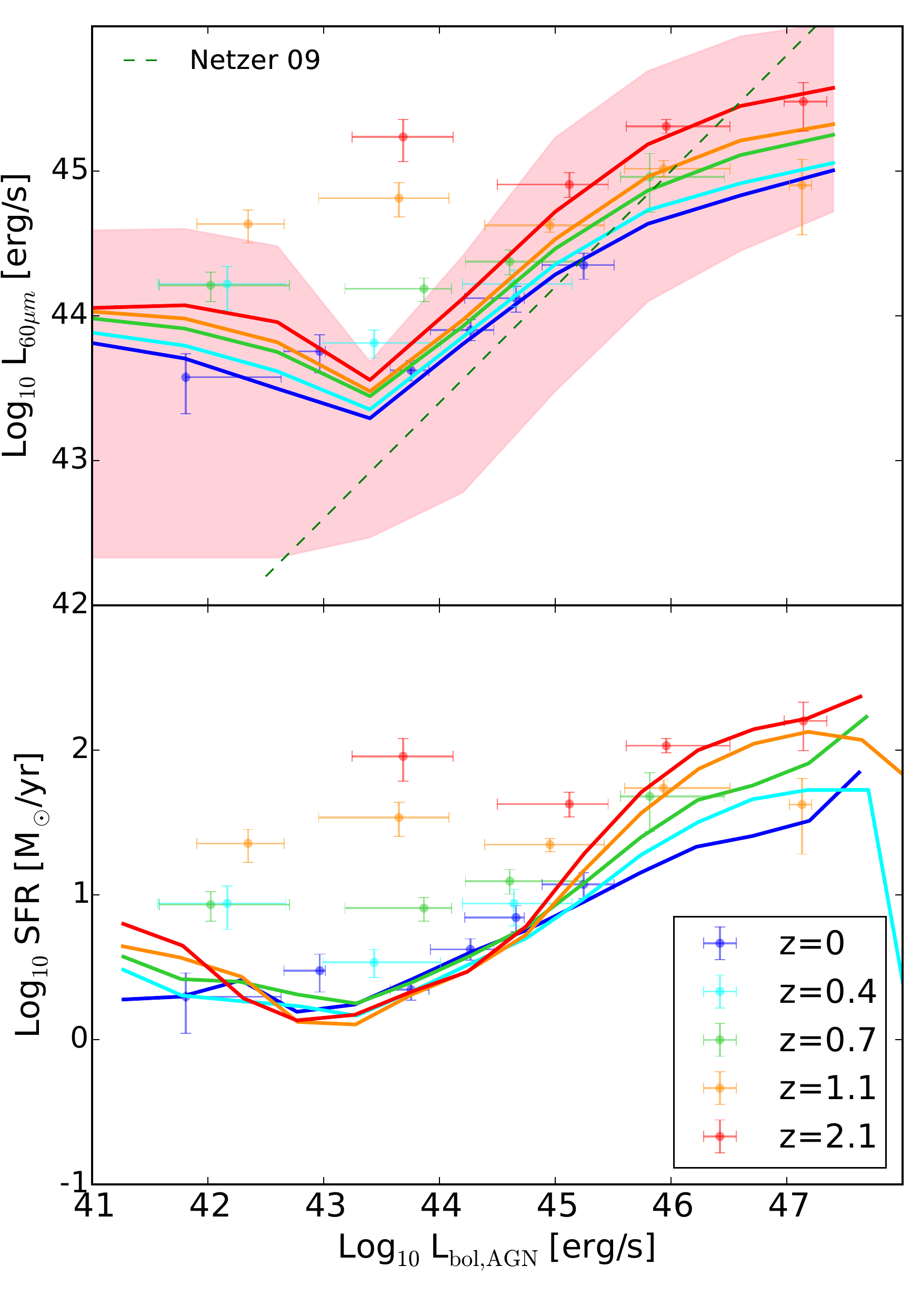}
  \caption{\emph{Top:} Average FIR emission at $60\micron$ as a function
    of AGN bolometric luminosity, $\Lbol$, for all galaxies in our
    model. Different lines indicate different redshifts as indicated
    by the key. {The shaded area represents the 10-90 percentiles for the model at z=2.1 (pink), but percentiles are similar at all redshifts.} Points with error bars show the observed
    $\Lsixty-\Lbol$ correlation in Rosario \etal (2012), where the $z=0$ data are from Swift-BAT. The
    solid-dashed line represents the correlation line connecting
    various AGN datasets (SDSS type-II AGN with $0.1\leq z\leq0.7$ and $\Lbol\gtrsim10^{42}\ergsec$, and Spitzer/IRS type-I quasars at $z\sim0.1$ and $z=2-3$) on the ${\rm SFR}-\Lbol$ plane in Netzer
    (2009). \emph{Bottom:} The predicted SFR-$\Lbol$ correlation as solid lines at different
    redshifts. We convert the Rosario et al. data points into SFRs according
    to Neistein \& Netzer (2014).}
  \label{fig:Lbol_sfrL60} 
\end{figure}

To compare our predictions with observational data, we show in Fig.~\ref{fig:Lbol_sfrL60} the average $\Lsixty$ vs. $\Lagn$ relation now for the whole galaxy sample and calculated in five different redshifts bins (solid lines). For the calculation of the average $\Lsixty$ we use the complete sample of galaxies at a given redshift without imposing any FIR luminosity cut. Our predictions are compared with the observational results from the PACS Evolutionary Probe (PEP) Herschel survey for the mean FIR emission of X-ray selected AGN at $0.4\leq z\leq2.1$ \citep{Rosario2012}. Also shown is the average FIR emission of local ($z=0$) X-ray selected AGN from the Swift-BAT sample \citep{Cusumano2010}. The straight dashed line has a slope of $0.8$ ($\Lsixty\propto\Lagn^{0.8}$) and shows the correlation line connecting various observational datasets on the ${\rm SFR}-\Lagn$ plane in \citealt{Netzer2009}. These datasets include type-II SDSS AGN ($0.1\leq z\leq0.7$ and $\Lbol\gtrsim10^{42}\ergsec$) and Spitzer type-I quasars at $z\sim1$ \citep{Netzer2007} and $z=2-3$ \citep{Lutz2008}. Interestingly, the predictions of $\galform$ for the $\Lsixty-\Lagn$ correlation show the same behaviour at all redshifts: for AGN with bolometric luminosities below $10^{43}$ ergs/s there is a negative correlation between SF and $\Lagn$, while the trend is reversed (e.g. positive correlation) for higher luminosities. 

Our model agrees well with the Swift-BAT data at $z=0$. In addition, the predicted trend for high luminosity AGN, i.e. $\Lagn\gtrsim10^{43}\ergsec$, follows the slope of the correlation line by Netzer, although the model flattens mildly in the regime of the brightest quasars. Moreover, the $\galform$ results at $\Lagn\gtrsim10^{44}\ergsec$ follow the observational data from Rosario \etal. {Both \citealt{Rosario2012} and \citealt{Stanley2015} note an increase in \Lsixty~at higher \Lagn, albeit the increase is weaker in \citealt{Stanley2015}. Our model does predict a moderate increase of SFR towards higher \Lagn. In our case, this is due to the fact that the high AGN luminosity end is dominated by the starburst mode where a strong correlation between SFR and \Lagn~is assumed (see section \ref{sec:twomodes}).}

 {However, at lower luminosities and higher redshifts the model diverges from observations. There is a pronounced dip in the model at around $\Lagn \approx 10^{43}\ergsec$, apparently where the model's two modes coincide (see figure \ref{fig:scatter_plot}), potentially an artifact of the abrupt transition between the two modes.} The negative trend at low luminosities is not visible at all in the data from Rosario \etal (2012), which are practically flat at all redshifts. This observed flat correlation was used in previous studies (\citeauthor{Rosario2012}; \citeauthor{Mullaney2012a}) as a hint of a possible disjoint evolution of BHs and their host galaxies, contrary to what is commonly assumed in many models of galaxy formation. Here we find that in $\galform$, this lower luminosity regime is purely shaped by the strong correlation between AGN activity and galaxy evolution.

We also show in the same plot (lower panel) the correlation between AGN luminosity and average SFR for the same sample of galaxies shown in the top panel. The Herschel observational data have now been divided by $1.9\times10^{43}\ergsec$ to convert the observed the $60\micron$ flux into a rough estimation for the SFR \citep{Neistein2014}. The model predictions show a behaviour similar to the $\Lsixty-\Lagn$ correlation shown in the top panel, though the disagreement with the observational data is now more evident. We note however, that our estimate of the observed SFR is only a crude approximation, and therefore this comparison with the model predictions is only for illustrative purposes.

\begin{figure}
  \centering
 \includegraphics[width=0.47\textwidth]{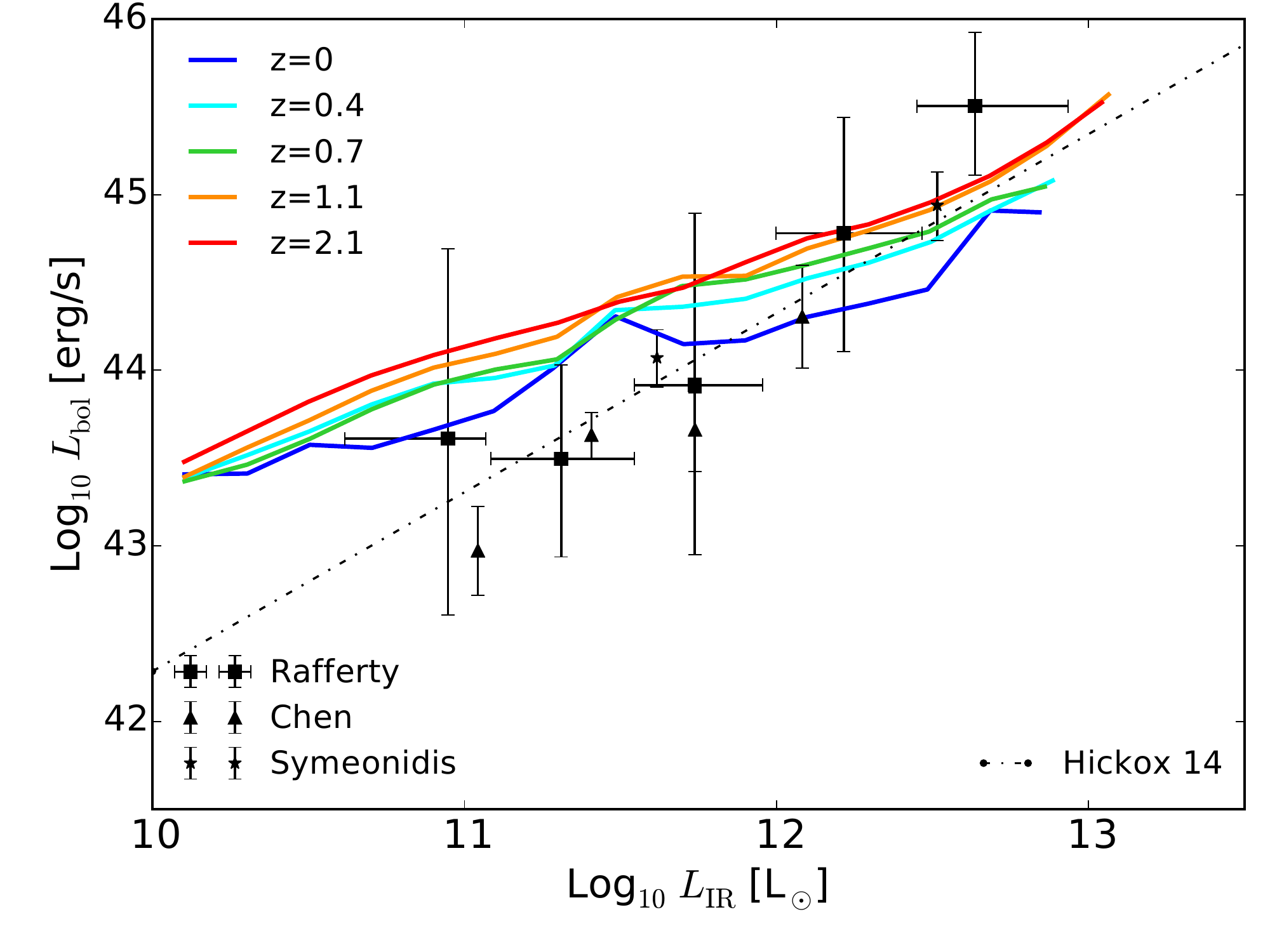}
  \caption{Average AGN bolometric luminosity, $\Lbol$, as a function of $\Lir$ ($8-1000\micron$), to allow a direct comparison with the observational data. Different redshifts are shown by different colours, as indicated by the key. Points with error bars show the observational data by Rafferty \etal (2011), Chen \etal (2013), and Symeonidis \etal (2011). The dot-dashed line represents the predictions of the theoretical model presented in Hickox \etal (2014).}
  \label{fig:Lir_Lbol}
\end{figure}

Finally, we plot in Fig.~\ref{fig:Lir_Lbol} the average AGN luminosity $\Lagn$ as a function of the total FIR luminosity $\Lir$. Our predictions are compared to recent observational data (\citealt{Rafferty2011}; \citealt{Symeonidis2011} and \citealt{Chen2013}) and the AGN variability model proposed by \citealt{Hickox2014}. Our model predictions are in moderate agreement with the observational results, as well as with the model proposed by Hickox \etal (even though the SFR and AGN activity are fully and directly coupled in $\galform$ without any AGN variability). The $\Lir-\Lagn$ correlation predicted by the model is entirely shaped by the starburst mode, ({this is in agreement with the observed correlation between SMBH growth rate and the SFR reported by \citealt{Mullaney2012b}}). In contrast to the predictions shown in Fig.~\ref{fig:Lbol_sfrL60}, we now find a monotonically increasing correlation. The average $\Lagn$ luminosity is dominated by those galaxies undergoing starburst-mode AGN activity, and thus the slope of the correlation reflects the slope of the starburst-mode branch in Fig.~\ref{fig:scatter_plot}. This is because the typical AGN luminosities of these galaxies, and also their number density, are much higher than the ones of galaxies in the hot-halo mode branch.

\subsection{Properties of AGN hosts}

The $\Lagn-\Lir$ correlation can in principle provide insights into the host properties of AGN and thus, impose constraints on theoretical models of galaxy formation. For example \citeauthor{Rosario2012} suggest that the flatness of the correlation, \Lagn-\Lsixty, at low AGN luminosities and its steep evolution at higher luminosities indicates two regimes of AGN activity. What our results show is that these two AGN regimes happen in very different galaxy populations.

AGN at the bright end of the $\Lagn-\Lsixty$ correlation shown in Fig.~\ref{fig:Lbol_sfrL60} are found in bursty systems experiencing intense SF. The hosts of these luminous AGN are gas rich systems living in $\sim 10^{12} M_{\odot}$ DM halos, and are similar to actively star forming galaxies. In contrast, in the faint $\Lagn$ region of the plane we find galaxies whose central BHs accrete predominately via the hot-halo mode. As already mentioned, this mode is particularly prominent in more massive systems ($\gtrsim 10^{13} \Msun$), where AGN feedback efficiently suppresses gas cooling and SF. The typical host of an AGN in that region would be a spheroidal system that at low redshifts resembles an early-type galaxy. 

\begin{figure}
  \centering
 \includegraphics[width=0.47\textwidth]{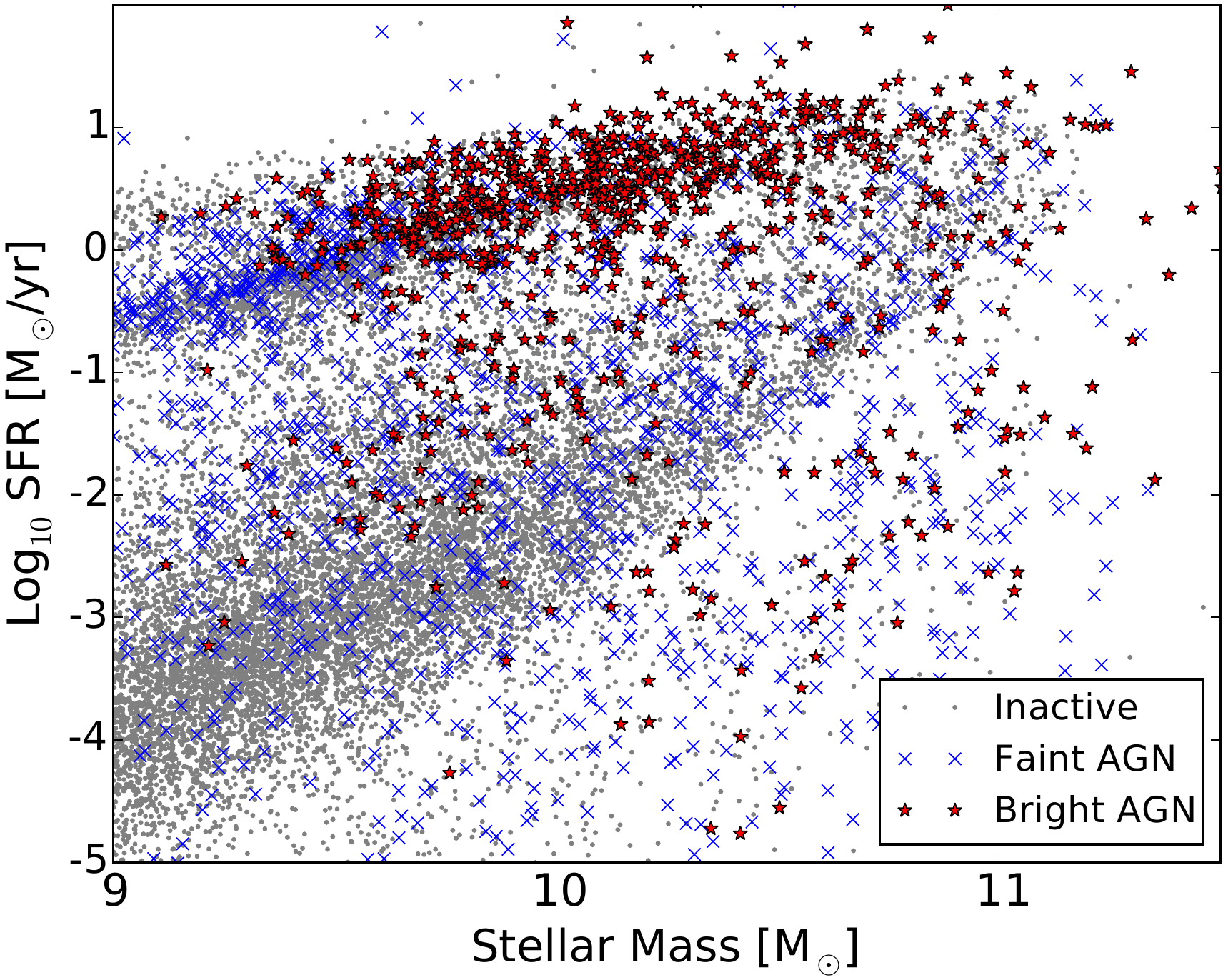}
  \caption{Stellar mass - star formation rate plane at $z=0$ for an unbiased subset of model galaxies in three bins: \emph{inactive:} $\Lagn<10^{41}$ erg/s, \emph{faint AGN:} $10^{41}<\Lagn<10^{43}$ erg/s, \emph{bright AGN:} $\Lagn>10^{43}$ erg/s }
  \label{fig:mstar-sfr} 
\end{figure}

{The stellar masses of the galaxies populating the $\Lagn-\Lir$ plane span a wide range of values (see Fanidakis et al. 2013b). We show this in Fig.~\ref{fig:mstar-sfr}, where we plot the SFR as a function of stellar mass at $z=0$ for active and inactive galaxies in our model. We define as inactive all galaxies with $\Lagn<10^{41}$ erg/s. At the same time, we split active galaxies into faint ($10^{41}<\Lagn<10^{43}$ erg/s) and bright AGN ($\Lagn>10^{43}$ erg/s). The transition luminosity from faint to bright AGN approximately marks the break in the $\Lagn-\Lsixty$ correlation shown in Fig.~\ref{fig:Lbol_sfrL60}. We refer the reader to \citet{Lagos2011} for the overall properties of galaxies on the $M_{\rm star}-SFR$ plane in $\galform$.}

{What is immediately evident from Fig.~\ref{fig:mstar-sfr} is that AGN and inactive galaxies populate the same regions on the $M_{\rm star}-SFR$ plane. Bright AGN are predominately found on the main sequence, though the scatter is strong. AGN in this sample are powered by accretion during the starburst mode. On the other hand, faint AGN are found both on the main sequence, but also immediately below it, i.e. in the region of passive galaxies. AGN hosts in the passive regime are typically quenched and have low SFRs. In the faint AGN sample we find AGN powered by both the starburst (AGN in the main sequence) and the hot-halo mode (AGN in the  passive regime). Overall, we find no particular feature, e.g. green valley, in the distribution of AGN. A detailed comparison between active and inactive galaxies could potentially reveal differences in the two galaxy populations, but this exercise is beyond the scope of this study. We finally note that the $M_{\rm star}-SFR$ plane at higher redshifts, ($z\sim2$) is characterised by the absence of the passive regime (for both active and inactive galaxies). The bimodal distribution of AGN at $z=0$ is linked to the quenching of galaxies in the low-redshift universe and relates strongly to the colour bimodality of AGN in $\galform$, as shown in \citet{Georgakakis2014}.}

\begin{figure*}
  \centering
  \includegraphics[width=0.67\textwidth]{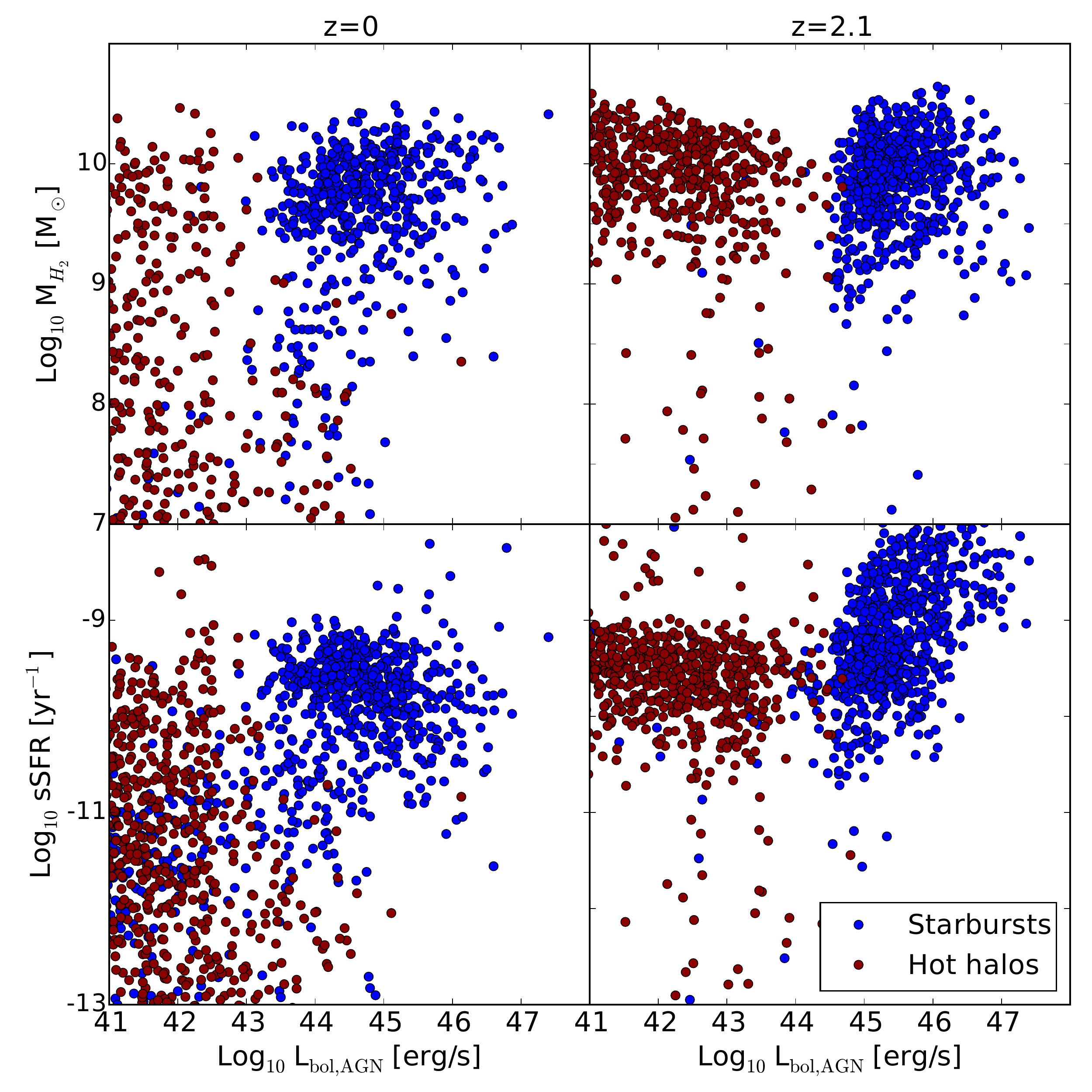}
  \caption{\emph{Top:} Predicted molecular hydrogen mass, $M_{H_2}$, as a function of bolometric
    AGN luminosity for $4,000$ randomly selected galaxies at $z=0$ (left) and
    $z=2.1$ (right). Galaxies
    in which the central BH experiences starburst-mode activity are shown in blue, while those with BHs growing during the hot-halo mode in red. \emph{Bottom:} Predicted specific SFR (SFR/M$_*$) as a
    function of AGN luminosity. Colors and redshifts as at the top panel. We
    only show galaxies with stellar masses above
    $10^{10}\Msun$.}
  \label{fig:coldgas}
\end{figure*}

To gain more insight into the SF properties of AGN hosts, we show in the top panels of Fig.~\ref{fig:coldgas}, the molecular hydrogen gas content as a function of AGN luminosity for a small subset of galaxies at $z=0$ and $z=2.1$. We remind the reader that the molecular hydrogen content is calculated following \citet{Lagos2011}, where the molecular-to-total gas ratio increases with the mid-plane hydrostatic pressure of the galactic disk. What is immediately evident in this plot is that both at high and low redshifts the hosts of faint and bright AGN reach similarly high molecular gas contents, but with a larger scatter towards lower values in faint AGN hosts at $z=0$. At a first look, this is surprising given that the hosts of faint AGN tend to be more passive systems. Indeed, the total cold gas reservoirs of early-type galaxies are lower than those of starbursts in our model. Yet, the model predicts a strong correlation between the molecular-to-atomic hydrogen ratio with increasing bulge-to-total ratio, meaning that early type galaxies are relatively richer in molecular hydrogen compared to atomic hydrogen, as implied also by past and recent observations \citep{Young1989,Bettoni2003,Leroy2008,Lisenfeld2011,Boselli2014}. This is due to the higher compactness of early-type galaxies compared to starburst galaxies, which results in higher gas pressure and thus, higher molecular-to-atomic hydrogen ratios \citep{Lagos2014b}. 

Despite the high $\mathrm{H}_2$ mass in early-type galaxies, the SFR in these systems remains low compared to starburst galaxies, as has already been shown in the lower panel of Fig.~\ref{fig:Lbol_sfrL60}. At high redshift this is due to the different timescales for star formation between spheroidals and starburst systems, with the latter ones having shorter timescales (see also section \ref{sfr_laws}). At low redshifts, early-type galaxies, \ie bulge-dominated systems, tend to have relatively low mean SFRs due to their lower mean molecular gas contents. This is also shown in the lower panel of Fig.~\ref{fig:coldgas}, where we now plot the specific SFR (sSFR).

Similarly to the SFR, the AGN luminosity is also low in early-type galaxies. This is due to the fact that these systems are dynamically stable, having very subdominant disks. They rarely undergo disk instabilities and thus, quasar-like AGN activity is never achieved (though major mergers could occasionally produce significant accretion). BHs typically grow via the hot-halo mode, which in the redshift regime of interest always produces low density accretion flows, irrespective of the cold gas reservoir of the galaxy. In contrast, in the high-luminosity regime we find gas-rich starburst galaxies that often experience disk instabilities which then trigger efficient and quasar-like accretion onto their central BHs.  

{Our results suggest that galaxies in the low luminosity region of the $\Lagn-\Lsixty$ plane are less efficient \emph{both} at building stars and growing BHs than their high luminosity counterparts.}

\section{Discussion}

Overall our model produces a SFR-\Lagn relation that is in reasonable agreement with observations, especially taking into account that no parameters of the model were tuned to reproduce this specific observational relation.
As it can been seen in Fig.~\ref{fig:scatter_plot}, the model predicts FIR luminosities for the faint and moderate luminosity AGN that match those observed in the Herschel surveys. In addition, the model provides a reasonably good fit to the FIR luminosity function of the total galaxy (Lacey et al. in prep) and AGN population \citep{Fanidakis2012}. Note that \citeauthor{Rosario2012} use stacking to estimate the FIR flux for sources whose flux is below the $3\sigma$ detection limit of the PACS instrument. Therefore, there is in principle no reason to create a mock catalogue for mimicking flux limits and biases related to the observations. Thus, assuming that their data sample is complete, possible reasons for the disagreement could be the following. First of all, there could be a strong contribution in the model of very faint FIR sources to the average $\Lsixty$ value. 

As we have already mentioned, the objects populating this region of the plot are identified with early type galaxies. These systems are passive in terms of their efficiency in forming of stars (although occasionally a starburst could occur as a consequence of a major merger), and therefore their SFR and, as a consequence, their FIR emission depend on the H$_2$ gas reservoir \citep{Lagos2011}. However, the model seems to produce a few too many low mass systems that are also low in cold gas (atomic and molecular) and thus have low SFR. But based on previous work with this model \citep[also Lacey et al. in prep.]{Lagos2012,Lagos2014a,Lagos2014b}, the FIR luminosities and SFRs in the vertical axes of Fig.~\ref{fig:Lbol_sfrL60} should be in principle consistent with the observations.

A more plausible reason for the strong disagreement could be the simplicity of the assumptions for the triggering and variability of AGN activity. In $\galform$, AGN activity begins with the onset of star formation in a starburst without a time delay. A possible delay between the triggering of AGN activity and the formation of stars in a burst could result into a decrease towards lower AGN luminosities for the objects populating the bright AGN regime of the diagram. The incorporation of an AGN variability model could have a similar effect. AGN luminosities are calculated assuming a constant accretion rate. Thus, the luminosity is constant during the entire course of accretion. However, AGN are known to exhibit strong variability on a wide range of timescales (\citealt{Novak2011}). 

{Recently, Hickox et al. presented  a model in which SF and BH accretion are closely connected over long timescales, but that short-term AGN variability can wash out this correlation for low to moderate \Lagn.  Here, despite the fact that the $\galform$ intrinsic FIR-AGN relation is driven by completely different phenomena, the inclusion of AGN variability could improve the agreement with the observations simply by shifting some bright quasars to lower AGN luminosities.}

A final explanation for the disagreement with the observations at lower AGN luminosity could be that the AGN luminosities we calculate for the AGN in the hot-halo mode are underpredicted (see also discussion about the hot-halo AGN in Krumpe et al. in prep). Indeed, the sharp transition from the hot-halo mode to the starburst mode, which appears as a strong break at $\Lagn\sim10^{43}\ergsec$ in the $\Lagn-\Lsixty$ correlation, could possibly be smoothed out if the hot-halo AGN luminosities were systematically higher. In the current version of the model, the hot-halo luminosity is calculated directly from the cooling properties of the host DM halo, via the expression $\dot{M}=L_{\rm cool}/(\epsilon_{\rm kin} c^2)$. While $L_{\rm cool}$ is well defined in the model, the value of the efficiency parameter $\epsilon_{\rm kin}$ is loosely constrained, mainly by requiring the model to reproduce the BH mass function in the local Universe. By boosting the luminosity of hot-halo AGN we could in principle make the transition from one regime to the other smoother and thus obtain a better agreement with the data. We note however that our aim in this paper is merely to report what observable correlations are predicted by a model in which BH and SFR are strongly coupled. We made no adjustments to the model in preparation for doing this analysis. Nevertheless, it is interesting that a comparison with FIR observations of X-ray selected AGN could provide possible constraints in the modelling of AGN feedback in galaxy formation models. 

\section{Conclusions}
In the current paradigm of galaxy formation the triggering mechanisms of AGN activity are often responsible for initiating intense SF in the host galaxy. At the same time, it is widely accepted, yet unproven, that AGN activity is the main driver of the quenching of SF in massive galaxies (\citealt{DiMatteo2005}, \citealt{Monaco2005}). In both cases, a correlation between observable proxies for SF and AGN activity is somehow expected. However, several recent observational studies have shown that there is little \citep{Rosario2012} or practically no correlation \citep{Mullaney2012a, Stanley2015} between FIR luminosity and the AGN luminosity. These results have been seen as a challenge for current galaxy formation models.

In this study, we have reported the predictions of the semi-analytic model $\galform$ for the connection between SFR and AGN activity. $\galform$ calculates AGN properties by tracking the BH accretion rate and the evolution of BH mass and spin. BHs grow via cold gas accretion (starburst mode), usually triggered by disk instabilities or galaxy mergers, and by hot-gas accretion (hot-halo mode), typically in massive quasi-hydrostatic haloes that are subject to AGN feedback. During the course of accretion the \galform~code calculates various AGN properties, as for example the disk emission in different bands, by considering the dependence of the bolometric disk emission on the structure of the accretion disk. 

Galaxies in $\galform$ build stars in disks from molecular hydrogen. When a galaxy experiences a disk instability or galaxy merger all the available cold gas (atomic and molecular) turns into stars via a burst. The growth of BHs is strongly linked to the buildup of stars. During a burst of SF a fraction of the gas that turns into stars is accreted onto the BH (starburst mode). This creates a strong correlation between SFR and AGN luminosity. In galaxies found in more massive, quasi-hydrostatic haloes SF is less efficient. The accretion power couples with the cooling properties of the gas in the halo resulting in a suppression of gas cooling and a regulation of the SF. This gives raise to an anti-correlation between SFR and AGN activity. 

To compare our predictions to recent Herschel observations for the mean SFR of X-ray selected AGN we compute the properties of galaxies in the FIR. The FIR emission in our model is due to the reprocessing by dust of the incident stellar continuum radiation. The resulting emission at $60\;\micron$, $\Lsixty$, scales nearly linearly with SFR (Fig.~\ref{fig:sfr_Lir}), and thus represents a good tracer for the SFR. For the brightest AGN, $\galform$ predicts a strong correlation between $\Lbol$ and $\Lsixty$, which arises from the strong coupling of the BH accretion to starbursts. For faint and intermediate luminosity AGN, the model predicts a mildly negative correlation, which is a consequence of the negative feedback effect these AGN have on their host galaxies. When compared to the Herschel PACS data for X-ray selected AGN in the COSMOS and GOODS-S/N fields we find a very good agreement in the bright AGN luminosity regime. In the low-luminosity regime we find that the model systematically underpredicts the average $\Lsixty$.

Finally, we showed that the objects populating the bright and faint regimes of the $\Lbol-\Lsixty$ plane represent different classes of galaxies. In the high luminosity regime we find gas-rich disk galaxies, that recently underwent a disk instability, and now exhibit strong starbursts and prodigious BH growth and AGN activity. In the low luminosity regime, we find massive early-type galaxies that experience quiescent SF, BH growth and AGN activity. Interestingly, despite the fact that these galaxies are subject to AGN feedback, many reach relatively large masses of molecular hydrogen. In fact, we find that their H$_2$ reservoirs at high redshift are similar to the gas-rich disk galaxies in the high luminosity regime of the diagram. These galaxies, however, are inefficient in building stars as can be seen in their sSFRs. We find that their SFRs and FIR luminosities tend to be on average approximately one order of magnitude lower than the galaxies in the bright luminosity regime. Nevertheless, a large fraction reach high molecular hydrogen masses, so a large population of low AGN luminosity galaxies should have as bright CO emission as their high-AGN luminosity counterparts. 

Future observations of the molecular gas/CO emission in X-ray selected AGN will help to further disentangle the SFR-AGN correlation. Until then, we hope that this work will provide adequate motivation that AGN activity remains a viable solution to explain the origin of red-and-dead galaxies.

\section*{Acknowledgments} 

The authors would like to thank David Rosario for providing the observational data in Fig. 3. Also many thanks to Roberto Decarli for important guidance and discussions. TAG and AVM acknowledge funding by Sonderforschungsbereich SFB 881 “The Milky Way System” (subproject A1) of the German Research Foundation (DFG).

\bibliographystyle{mn}
\bibliography{bib}

\appendix

\bsp

\label{lastpage}

\end{document}